# Achromatic imaging systems with flat lenses enabled by deep learning

Roy Maman*, Eitan Mualem, Noa Mazurski, Jacob Engelberg and Uriel Levy*

Institute of Applied Physics, The Faculty of Science, The Center for Nanoscience and Nanotechnology, The Hebrew University of Jerusalem, Jerusalem 91904, Israel
E-mail: Roy.maman@mail.huji.ac.il, ulevy@mail.huji.ac.il

### Abstract

Motivated by their great potential to reduce the size, cost and weight, flat lenses, a category that includes diffractive lenses and metalenses, are rapidly emerging as key components with the potential to replace the traditional refractive optical elements in modern optical systems. Yet, the inherently strong chromatic aberration of these flat lenses is significantly impairing their performance in systems based on polychromatic illumination or passive ambient light illumination, stalling their widespread implementation. Hereby, we provide a promising solution and demonstrate high quality imaging based on flat lenses over the entire visible spectrum. Our approach is based on creating a novel dataset of color outdoor images taken with our flat lens and using this dataset to train a deep-learning model for chromatic aberrations correction. Based on this approach we show unprecedented imaging results not only in terms of qualitative measures but also in the quantitative terms of the PSNR and SSIM scores of the reconstructed images. The results pave the way for the implementation of flat lenses in advanced polychromatic imaging systems.

### 1. Introduction

In recent years, advances in optical technology have paved the way for innovative solutions that challenge traditional approaches to lens design. Among these breakthroughs, flat lenses(*1*), including diffractive lenses(*2–5*) and metalenses(*6–13*), have emerged as compelling alternatives. One of the major obstacles to widespread use of flat lenses is their inherent chromatic aberration(*14*), that is, the focal length of the flat lens is strongly dependent on the wavelength of illumination. Novel nanostructure design has introduced the ability to correct the chromatic aberration of metalenses(*15–22*) and of diffractive lenses(*5, 23–26*). Both have been shown to have fundamental performance limitations (*27, 28*). Another way to tackle the chromatic aberration is by image processing (*29*). Chromatic aberration caused by single-ball lenses has been corrected to some extent by an image processing algorithm (*30*). With the rapid advancement of machine learning, deep learning models have demonstrated superior performance compared to traditional image restoration methods(*31–33*). Recent works show a step forward towards facing flat lens challenges by applying algorithms to reduce chromatic aberrations of a hybrid diffractive-refractive imaging system using images that were captured indoors in laboratory conditions(*30, 34, 35*). Yet, the "holy grail" of eliminating chromatic aberration over the entire visible spectrum in imaging systems based on flat lenses operating outdoors in natural ambient light conditions remains an unsolved challenge. Our work tackles this grand challenge. Hereby, we have constructed a dual camera system that captures chromatic-aberrated images from a flat lens, alongside with aberration-free images that are captured by a conventional refractive lens. Using this system, we have created the first ever dataset of outdoor images taken with a flat lens over a wide field of view (FOV), along with their ground truth version. Having this dataset at hand, we trained a deep-learning model for the correction of chromatic aberrations. As a result of this, we were able to obtain high quality visual images over the entire visible spectrum based on natural outdoor conditions. On top of the high qualitative assessment, our approach shows excellent quantitative results in terms of PSNR and structure similarity index (SSIM).

## 2. Background/Related work
### 2.1 Chromatic Aberrations of Thin Lenses

The chromatic aberration of diffractive lenses and conventional metalenses is given by Eq. (1) (*14*)

$$\Delta f = f \frac{\Delta \lambda}{\lambda} \qquad (1)$$

where $\lambda$, and f are the nominal (central) wavelength and focal length, respectively, of the imaging system. $\Delta \lambda$ is the wavelength band over which the system operates, and $\Delta f$ is the corresponding change in focal length, i.e., the longitudinal chromatic aberration for an object at infinity. The transverse axial chromatic aberration (TAC) is then given by Eq. (2)(*14*)

$$TAC = \Delta f \frac{D}{2f} = \frac{\Delta \lambda}{\lambda} * \frac{D}{2} \qquad (2)$$

Where D is the lens aperture diameter. The geometrical point-spread-function (PSF), for a single wavelength, shifted by $\Delta \lambda$ from the nominal value, will be a top-hat profile with radius equal to this transverse aberration (neglecting diffraction effects). In the case of a thin lens with a large field-of-view, off-axis aberrations come into play. The lateral chromatic aberration of a thin lens is given by Eq. (10) (*14*).

$$TLC = TAC * \frac{y_p}{y} \qquad (3)$$

Where $y_p$ is the chief ray height at the lens, and y is marginal ray height at the lens. The effect of both types can be seen visually by comparing an image taken through a flat lens (Fig. 1, top row) and an image taken through a standard refractive lens (Fig. 1, bottom row). The image sensor is positioned at the focus of the flat lens for the center of the green spectrum. As can be seen, the TAC causes a slight defocus in the green channel, and a major defocus in the red and the blue channels. The TLC shifts each wavelength in the image plane, according to Eq. (3), and causes a radial spatial separation between the different colors, which is especially noticeable when observing sharp gradients in the image.

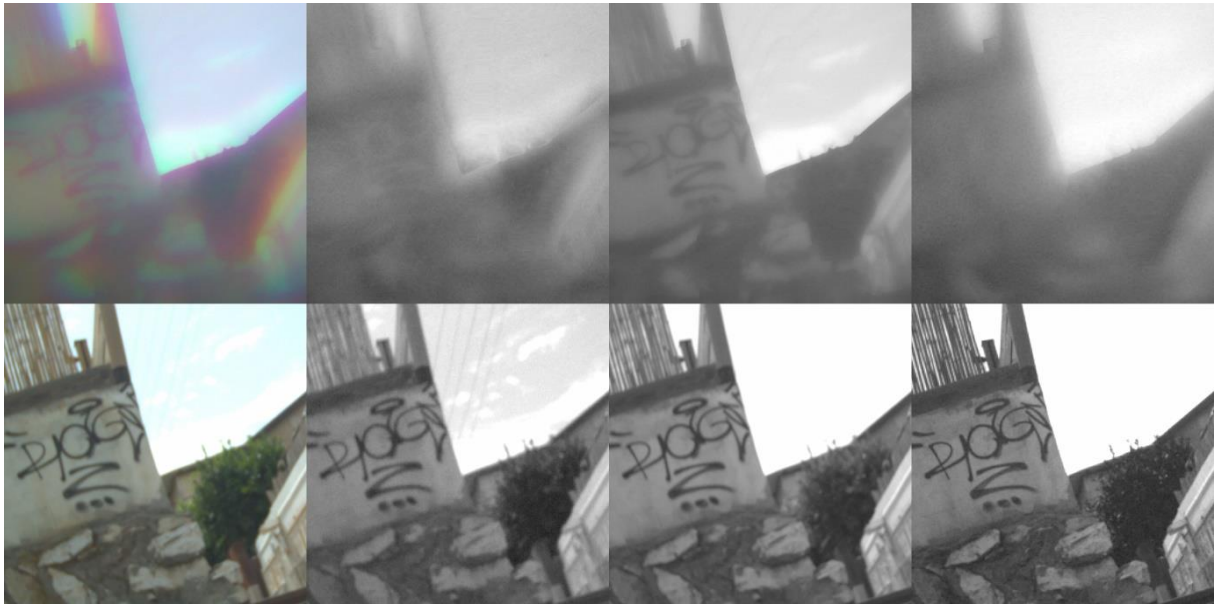

Figure 1 – **The effect of the TAC and the TLC on the different color channels. Top** – an aberrated off-axis section of an image captured with the flat lens (left) together with its 3 color channels (R, G, B). **Bottom** – the corresponding ground truth image, captured with the refractive lens (left) together with its 3 color channels (R, G, B).

## 2.2 Traditional Chromatic Aberrations Correction Methods

The effect of TAC is manifested as image blurring in each color channel. Most of the traditional methods for TAC correction use the lens PSF to deconvolve the image(*36*)(*37*)(*38*). Computing the exact PSF is a challenging task that requires in-depth knowledge of the lens structure. There are several algorithms for PSF estimation(*39–42*). TLC adds further geometric errors to the image. Deconvolution methods with TLC correction priors can correct TLC and TAC simultaneously. Suggested TLC correction priors use cross-channel gradients(*29*) or cross-channel shearlet transform(*30*). These traditional methods do not rely on ground truth images, and therefore cannot report metric scores on real images. Importantly, such methods become less efficient in low SNR scenarios, as they tend to enhance the noise.

### 2.3 Deep Learning-based Image Reconstruction

In recent years, convolutional neural networks(*43–46*) have demonstrated superior performance compared to traditional restoration methods(*31–33*). The Transformer model, first developed for natural language tasks(*47*), has been adapted in numerous vision tasks such as image recognition(*48–50*), and object detection(*51, 52*). Transformers (*48, 49*) break down an image into a sequence of patches and learn how these patches are interrelated. These models can learn long-range dependencies between image patch sequences but come with high computational complexity (*53*). With recent advances (*51, 54*), transformers were successfully applied to low-level vision problems such as defocus deblurring(*55, 56*) and image colorization(*57*).

## 3. Image Acquisition System and Process
### 3.1 Image Acquisition System

Motivated by the potential of deep learning algorithms to overcome chromatic aberrations of flat lenses, we turned to generating a comprehensive dataset. To capture chromatic-aberrated and sharp images simultaneously, we have constructed a dual camera system (Fig. 2). Our system consists of a beam splitter and two cameras so that each of the two cameras can capture the same scene. The beam splitter is suitable for the visible spectrum (Thorlabs CCM1-BS013/M) and is covered with an IR-cut filter. One camera uses a homemade diffractive lens while the other uses a commercial refractive lens (PT-0618MP f=6.0mm, F/1.8 Mega Pixel CCTV Board Lens). The two cameras are of the same model (Thorlabs CS165CU/M - Zelux® 1.6 MP Color CMOS Camera). The cameras are connected to a laptop and controlled by Thorcam TSI overlay plugin that allows sampling from two cameras simultaneously (up to a <1ms jitter).

As a flat lens we have used a simple binary diffractive lens, i.e., consisting of two phase levels. Such a lens is limited in each diffraction efficiency to ~ 40%, but nevertheless, has the same resolution and chromatic aberration as a multi-phase-level flat lens, be it a diffractive lens or a metalens. Therefore, we can use this lens to test our algorithm. Our flat lens is manufactured using e-beam lithography (Elionix ELS-G100), in PMMA resist, on a glass substrate. It has a quadratic phase function, with focal length of 5.2mm at 550nm. A mechanical stop is placed at the front focal plane of the lens, with diameter of 1.5mm (F/3.5). This yields correction of all monochromatic aberrations (except distortion which does not affect resolution), leaving us with the problem of correcting the chromatic aberrations described in section 2.1 (*14*).

The FOV of our system is limited by the mechanical aperture of the beam-splitter, and in addition we crop the image to speed-up our algorithm (our images use 512 × 512 pixels out of 1440 x 1080). Considering the focal length of 5.2mm and pixel size of our camera (3.45µm) this corresponds to an angular FOV of ~±10° in both horizontal and vertical dimensions (this is based on the characteristic lens distortion for this type of landscape lens system, which causes a sine relation between the angle and image height(*4*)). Measured data for our lens, including physical height profile and simulated MTFs on and off-axis, are given in the supplementary.

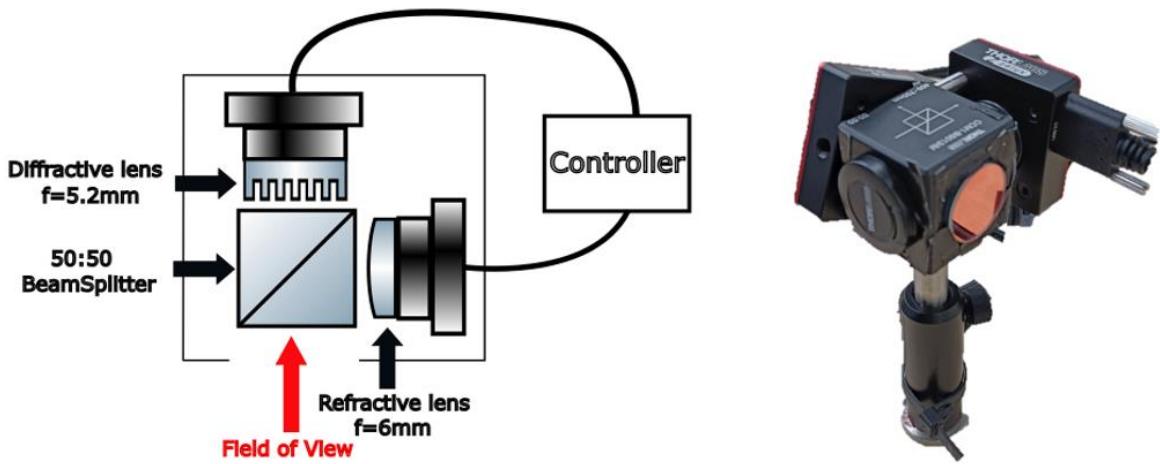

**Figure 2** – **Image Acquisition System. Left** – optical scheme. The beamsplitter splits the light equally between the diffractive lens and the refractive lens. The two lenses are connected to a camera of the same type and two images are sampled simultaneously by the controller. **Right** – The actual system.

### 3.2 Image Acquisition Process

To sample many different images, our system is installed on a vehicle, and images are taken automatically every 1 second. To avoid motion blur, the shutter speed is set to 0.4ms for the refractive-lens camera and 1ms for the diffractive-lens camera (since the flat lens F-number is higher than that of the refractive lens, we increased the exposure time for this camera to get the same average value as the other camera). Driving speed is lower than 10km/h. We captured all images in raw format.

### 4. Postprocessing

The generation of a dataset consisting of ground truth images that are perfectly aligned and identical in content with the images taken by the flat lenses poses several challenges. For the sharp image to be used as the ground truth for the aberrated image, both images must show the exact same field. Yet, in our sampling setup (Fig. 2), the two cameras have different focal lengths (5.2mm for the flat lens and 6mm for the refractive lens) and therefore different effective magnifications. The cameras are not optimally aligned in front of the beam-splitter, with one rotated relative to the other, because of mechanical constraints. Also, the sharp images are flipped by the beam-splitter and the field of view of both cameras is limited by it's mechanical aperture, so the sampled images must be cropped (Fig. 3).

In addition, the transformation needed to align the sharp image with the aberrated image is not constant for the entire dataset of images. This is due to the inherent time jitter in the image sampling, and the constant movement of the vehicle. To address the accumulated misalignment between the images, it is necessary to evaluate the optical flow(*58*). TV-L1 is a popular algorithm for optical flow estimation introduced by Zach(*59*) et al., improved in (*60*) and detailed in (*61*). In our postprocessing procedure, we use TV-L1 to estimate the optical flow between the green channel of the aberrated image and the green channel of its corresponding flipped sharp image. We then use the output of TV-L1 as a coordinate transformation for the entire aberrated image. Both images (the transformed aberrated image and the flipped sharp image) are cropped around the center to a size of 512x512 pixels (Fig. 3). The optical flow is estimated separately for each image pair. As the TV-L1 computes the flow for each pixel, the described transformation is elastic (non-rigid) and may add stitching artifacts. To overcome these artifacts to some extent, we went through all the images manually, and removed images with significant artifacts.

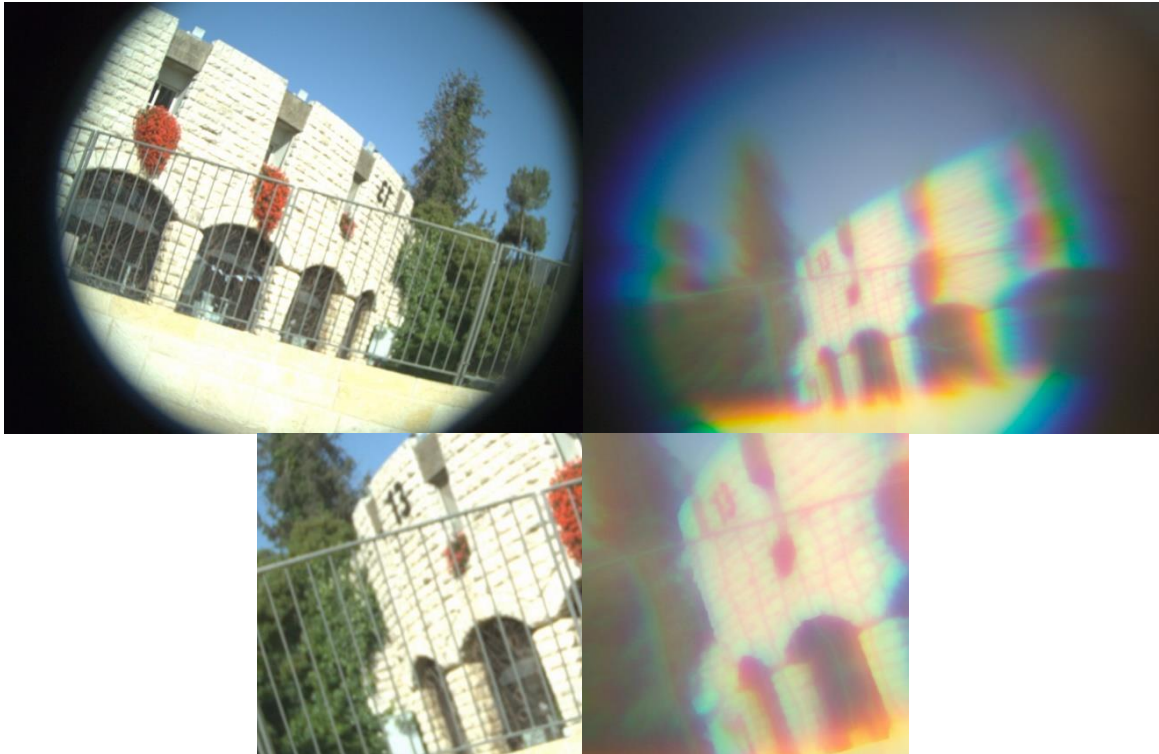

**Figure 3** – **An aberrated image (right) with its ground truth image (left). TOP –** Full images as captured by the acquisition system. **Bottom –** Cropped images after the postprocessing alignment.

## 5. Experiments
### 5.1 The Dataset

We used our image acquisition system to sample images with a diffractive lens together with ground truth images with a refractive lens. The images were later aligned using our postprocessing scheme. The dataset contains a total of 1414 image pairs of size 512x512 pixels. The datasets were then randomly divided into training (1131 images) and testing (283 images) subsets. The datasets are available online at:

*https://drive.google.com/file/d/12flN2HAXqMOcIVzjztS_15gvtkKeULOn/view?usp=sharing*

### 5.2 Network Architecture

We used the Restormer architecture (*54*), an efficient transformer model that was successfully applied for defocus deblurring. The primary computational burden in transformers arises from the self-attention (SA) layer(*47, 62*), whose time and memory complexity grows quadratically with image size. To make SA feasible for image restoration tasks, the Restormer introduces MultiDconv Head Transposed Attention (MDTA), which has a linear complexity. MDTA performs query-key feature interaction across channels rather than the spatial dimension. To transform features, the regular feed-forward network (FN) (*47, 62*) operates on each pixel location separately and identically. To improve representation learning, the Restormer proposes the Gated-Dconv Feed-Forward Network (GDFN). The GDFN allows each level of the module to focus on fine details complimentary to other levels. Like the regular FN, the GDFN uses a convolution layer to expand the feature channels. Our Restormer employs a 4-level encoder-decoder. From level-1 to level-4, the number of Transformer blocks are (4; 6; 6; 8), attention heads in MDTA are (1; 2; 4; 8), and number of channels are (48; 96; 192; 384). The refinement stage contains 4 blocks. The channel expansion factor in GDFN is =2:66.

### 5.3 Training Process

We train the model with AdamW optimizer ($\beta 1 = 0.9$, $\beta 2 = 0.999$, weight decay of $1e^{-4}$) for 300K iterations with initial learning rate of $3e^{-4}$ gradually reduced to $1e^{-6}$ with the cosine annealing(*63*). For progressive learning, we start training with patch size of 64x64 and batch size of 32. The patch size and batch size pairs are updated to $((64^2, 20), (80^2, 16), (96^2, 8), (128^2, 8), (160^2, 4))$ at iterations (92K, 156K, 204K, 240K, 276K). We used the following loss function to get the best results for the two common metrics, PSNR and SSIM:

$$L(gt, im) = \alpha MSE(gt, im) + \beta\big(1 - SSIM(gt, im)\big) \qquad (4)$$

Where $gt, im$ are the ground truth image and the reconstructed image, respectively. MSE is the mean square error function and SSIM is the structural similarity index metric(*64*).

### 6. Results

Figure 4 shows the training convergence curve of the described model, over 300K iterations. As can be seen, the loss function converges while the SSIM score is still increasing monotonically.

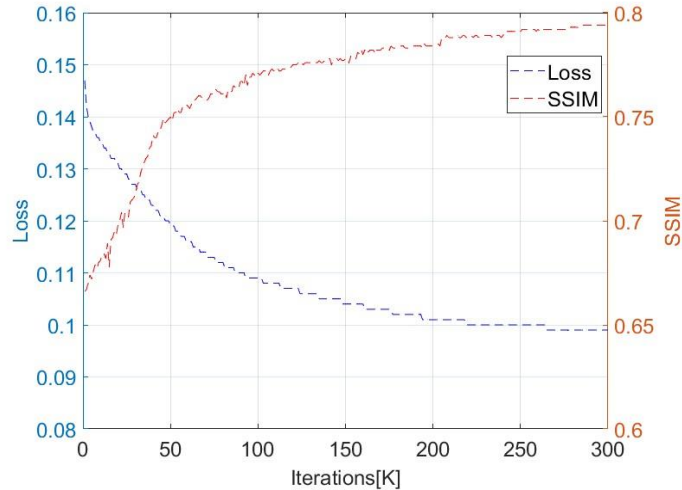

**Figure 4 – Training graph.** The loss and the SSIM scores as measured on the test set every 1k iterations.

We calculate both PSNR and SSIM scores for color images. Over the 283 test images, the average PSNR and SSIM are 30.82 and 0.78 respectively. Figure 5 shows the aberrated (top row), the ground truth (middle row) and the reconstructed images (bottom row) for 5 selected image pairs from the test set. It is clear to the naked eye that the images are reproduced sharply and accurately, albeit with some (small) inevitable loss at the high spatial frequencies. While some of the colors in the reconstructed images are a bit faded compared to the original image, the overall reconstruction quality is high and very convincing.

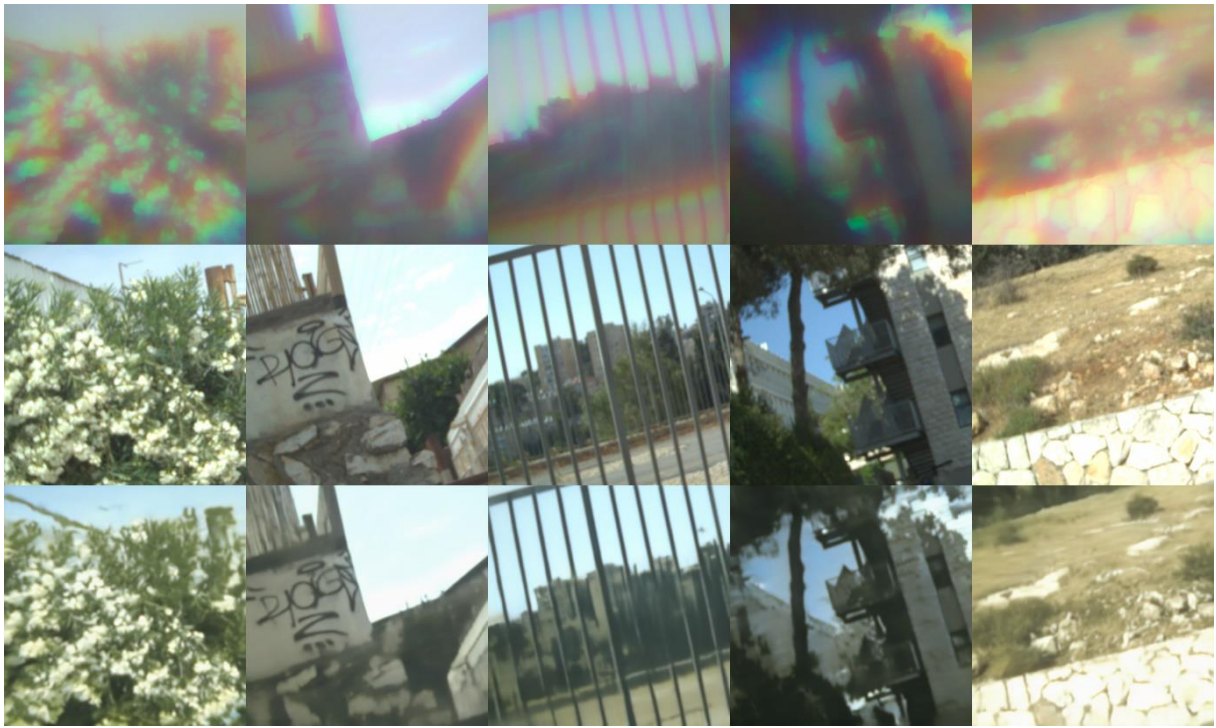

**Figure 5 – Visual results. Top row –** 5 aberrated test-set images. **Middle row –** The corresponding ground truth images. **Bottom row –** The reconstructed images.

Several existing methods calculate the PSNR/SSIM score using the Y channel in YCbCr color space(*54*, *65*). By considering the Y channel only in our set of images, the effect of the faded colors on the SSIM score is eliminated. The average SSIM score in this case is as high as 0.88. As our result is, to the best of our knowledge, the first model trained to correct chromatic aberrations from a flat diffractive lens, we could not compare it to a flat lens benchmark. Therefore, we have decided to adopt a more challenging evaluation criterion by comparing our results with those achieved by applying deblurring algorithms for chromatic aberrations caused by other lenses with significantly lower chromatic aberrations. In (*34*, *35*) the U-net architecture(*66*) was used on a small dataset of images captured from a high-resolution monitor using a Hybrid refractive-diffractive lens system. Although the chromatic aberrations in a hybrid lens system are much smaller than in our case, the reported PSNR is 27.71, significantly lower than our result. SSIM was not reported.

## 7. Conclusion

We propose a deep-learning-based approach to correcting chromatic aberrations originating from conventional flat lenses and use it to overcome chromatic aberrations of flat lens-based imaging systems operating outdoors under normal ambient illumination conditions. Qualitatively, our approach produced reconstructed images with excellent quality. Quantitatively, we were able to obtain image criteria of PSNR and SSIM that by far exceed those reported in previous works for chromatic aberrations correction. Our method deals with both TLC and TAC without the need for a PSF estimation. Finally, we offer a dataset of real outdoor images captured through a flat lens, to allow the community to further improve image reconstruction of flat lenses for color imaging. Our image acquisition system and process may be implemented for other lenses as well. We believe that the demonstrated results will pave the way for the widespread implementation of flat lenses in diverse applications benefiting from reduction in size, cost, and weight.

**Acknowledgements**

The flat lens was fabricated at the center for nanoscience and nanotechnology of the Hebrew University. The research was partially supported by the Israeli innovation authority.

# Achromatic imaging systems with flat lenses enabled by deep learning

*Roy Maman\*, Eitan Mualem, Noa Mazurski, Jacob Engelberg and Uriel Levy\**

Institute of Applied Physics, The Faculty of Science, The Center for Nanoscience and Nanotechnology, The Hebrew University of Jerusalem, Jerusalem 91904, Israel
E-mail: Roy.maman@mail.huji.ac.il, ulevy@mail.huji.ac.il

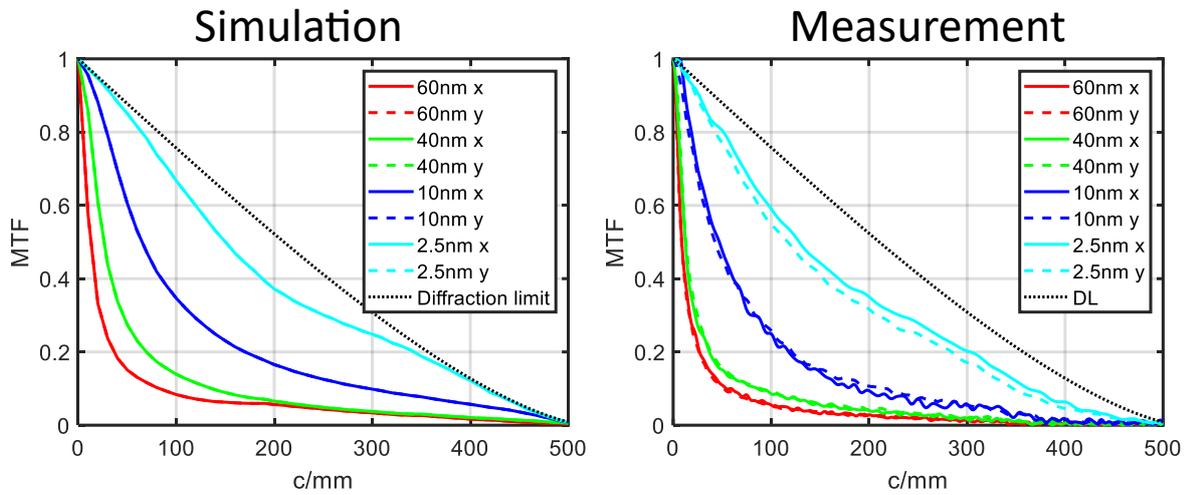

**Figure S1: MTF on-axis.** Simulation (left) and measurement (right) are shown for several spectral widths around 550nm. The 60nm spectral width most closely corresponds to the green channel of the camera used in our experiment.

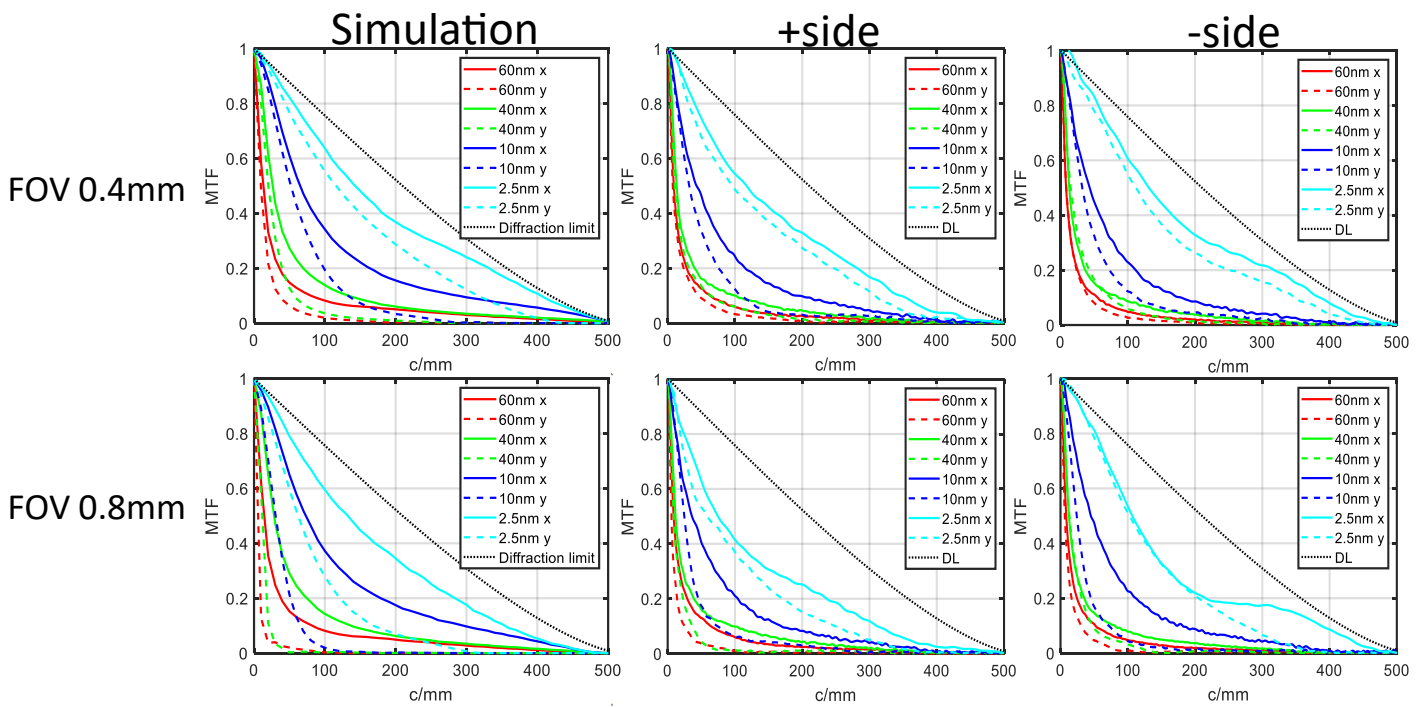

**Figure S2: MTF off-axis. Top –** FOV 0.4mm, simulation (left) and measurements on both lateral sides (right). **Bottom –** FOV 0.8mm, simulation(left) and measurements on both lateral sides (right). FOVs are in the image plane, and correspond to field angles of 4.4° and 8.85°.

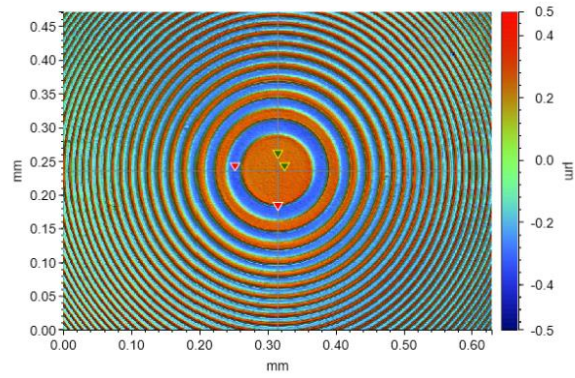
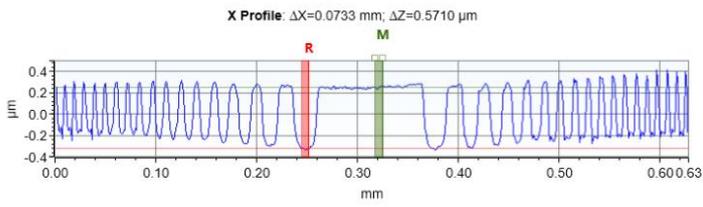 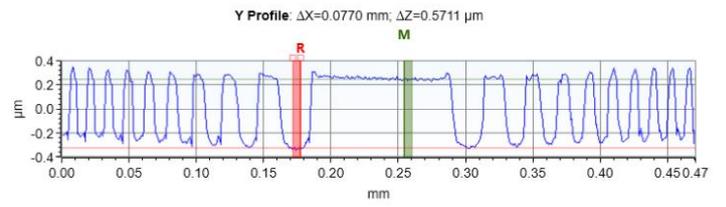

**Figure S3: Surface profile of the central part of our binary diffractive lens, measured using Bruker ContourGT-K optical profilometer.** The nominal step height was 558nm, corresponding to a phase of $\pi$, for PMMA refractive index of 1.49. The actual measured height can be seen in the profile to be 571nm.